\documentclass[a4paper]{jpconf}
\begin{document}
\title{Acceleration disturbances due to local gravity gradients in ASTROD I}

\author{Sachie Shiomi}
\address{Department of Physics, National Tsing Hua University,
Hsinchu, Taiwan 30013 R.O.C.} \ead{sshiomi@phys.nthu.edu.tw}

\begin{abstract}
The Astrodynamical Space Test of Relativity using Optical Devices
(ASTROD) mission consists of three spacecraft in separate solar
orbits and carries out laser interferometric ranging. ASTROD aims at
testing relativistic gravity, measuring the solar system and
detecting gravitational waves. Because of the larger arm length, the
sensitivity of ASTROD to gravitational waves is estimated to be
about 30 times better than Laser Interferometer Space Antenna (LISA)
in the frequency range lower than about 0.1 mHz. ASTROD I is a
simple version of ASTROD, employing one spacecraft in a solar orbit.
It is the first step for ASTROD and serves as a technology
demonstration mission for ASTROD. In addition, several scientific
results are expected in the ASTROD I experiment. The required
acceleration noise level of ASTROD I is 10$^{-13}$ m s$^{-2}$ at the
frequency of 0.1 mHz. In this paper, we focus on local gravity
gradient noise that could be one of the largest acceleration
disturbances in the ASTROD I experiment. We have carried out
gravitational modelling for the current test-mass design and
simplified configurations of ASTROD I by using an analytical method
and the Monte Carlo method. Our analyses can be applied to figure
out the optimal designs of the test mass and the constructing
materials of the spacecraft, and the configuration of compensation
mass to reduce local gravity gradients.
\end{abstract}

\section{Introduction}
A gravitational mission, Astrodynamical Space Test of Relativity
using Optical Devices (ASTROD) \cite{ASTROD1, ASTROD2}, was proposed
to test relativistic gravity, measure the solar-system parameters
with high precision and detect gravitational waves from massive
black holes and galactic binary stars. The concept of ASTROD is to
put two spacecraft in separate solar orbits and carry out laser
interferometic ranging with a spacecraft near the L1/L2 point. A
simple version of ASTROD, ASTROD I, has been studied as the first
step to ASTROD. ASTROD I employs one spacecraft in a solar orbit and
carries out interferometric ranging and pulse ranging with ground
stations \cite{Ni2002}.

The acceleration disturbance goal of ASTROD I is 10$^{-13}$ m
s$^{-2}$ Hz$^{-1/2}$ at frequency $\nu$ of 0.1 mHz. Assuming a 10 ps
timing accuracy and the acceleration noise of $10^{-13}$ m s$^{-2}$
Hz$^{-1/2}$ at frequency of about 0.1 mHz, a simulation for 400 days
(350-750 days after launch) showed that ASTROD I could determine the
relativistic parameters $\gamma$ and $\beta$, and the solar
quadrupole parameter $J_2$ to levels of 10$^{-7}$, 10$^{-7}$ and
10$^{-8}$, respectively \cite{Tang2004}. In order to achieve the
acceleration disturbance goal, a drag-free control system using
capacitive sensors will be employed.

A preliminary overview of sources and magnitude of acceleration
disturbances for ASTROD I is given by Shiomi and Ni
\cite{Shiomi2005}. According to their estimates assuming simple
models, local gravity gradients can be one of the largest
contributions to acceleration disturbances in ASTROD I. Therefore,
an elaborate gravitational modelling seems necessary.

The sources of acceleration disturbances due to local gravity
gradients can be classified into two categories. One is thermal
deformation of the spacecraft and the payload, mainly caused by
solar radiation. Inherent temperature fluctuations in solar
radiation result in unwanted acceleration. Because composing
materials of the spacecraft and payload vary in thermal expansion
coefficient, there would be complicated relative positional changes
inside of the spacecraft. Elaborate thermal modelling works are
required for a complete analysis. However, we will not discuss this
in this paper. The other is positional fluctuations of the test
mass. Even when there is no deformation in the spacecraft and
payload, positional fluctuations of the test mass produce unwanted
acceleration.

Xu and Ni did a preliminary work of gravitational modelling for the
ASTROD mission \cite{Xu2003}. They calculated gravitational
interaction between a test mass and a cylindrical reference mass (a
hollow cylinder with two end disks) by using the expression derived
for the shape design of STEP (Satellite Test of the Equivalence
Principle \cite{STEP}) test-masses \cite{Lockerbie1993}. They
evaluated the magnitude of gravitational acceleration caused when
the test mass, located at the centre of the reference mass, was
shifted along the axial axis of the reference mass. The expression
they used was obtained for the analyses of cylindrical bodies with
homogeneous density. We use more general expressions (section
\ref{st:expression of Fx}), applicable to arbitrary shapes, and
consider the gravitational acceleration between the ASTROD I test
mass and cylindrical bodies (section \ref{st:SC-TM}), and
rectangular parallelepiped objects (section \ref{st:RP-TM}) in this
paper. Also, we carry out the Monte Carlo simulation (section
\ref{st:MC}) to estimate the gravitational acceleration due to
positional fluctuations of the ASTROD I test mass.

\section{The configuration of the ASTROD I spacecraft}\label{configuration}

The ASTROD I spacecraft has a cylindrical shape with diameter 2.5 m
and height 2 m. Its surface is covered with solar panels. The
cylindrical axis is perpendicular to the orbit plane and a telescope
on board is set to point toward a ground laser station. The total
mass of the spacecraft is about 350 kg and that of payload is
100-120 kg (see \cite{Ni-Shiomi2004} and \cite{Ni2004} for more
detailed descriptions of the configuration). The orbit distance from
the Sun varies from about 0.5 AU to 1 AU (see figure 2 of
\cite{Ni-Shiomi2004} for a detailed description).

The test mass ($M_{TM}$ = 1.75 kg) is a rectangular parallelepiped
(50 $\times$ 50 $\times$ 35 mm$^3$) made from Au-Pt alloy with
density of 2 $\times$ 10$^4$ kg m$^{-3}$. The test mass is located
at the centre of the spacecraft. The six sides of the test mass are
surrounded by electrodes mounted on the housing for capacitive
sensing. The gap between each side of the test mass and the opposing
electrode is 2 mm.

\section{Acceleration of a test mass}\label{st:expression of Fx}

The gravitational potential energy of a test mass (density
distribution $\rho_t (\mathbf{x'})$ and volume $v_t$) in a
gravitational field produced by a source mass (density distribution
$\rho_s (\mathbf{x})$ and volume $v_s$) can be written by:
\begin{equation}
V = - 4 \pi G \sum_{l=0}^{\infty} \sum_{m=-l}^{l} \frac{1}{2l+1}
q_{lm} Q_{lm} \label{eq:V}
\end{equation}
where
\begin{eqnarray}
q_{lm} = \int_{v_t} \rho_t (\mathbf{x'})r'^{l}Y^{*}_{lm}(\theta',
\phi') d^3 x' \\ Q_{lm} = \int_{v_s} \rho_s
(\mathbf{x})r^{-(l+1)}Y_{lm}(\theta, \phi) d^3 x
\end{eqnarray}
Inner gravitational multipole moments and outer gravitational
multipole moments, $q_{lm}$ and $Q_{lm}$, represent the mass
distribution of the test mass and the source mass, respectively. $G$
(= 6.67 $\times$ 10$^{-11}$ N m$^2$ kg$^{-2}$) is the gravitational
constant.

The force between the test mass and the source mass in the sensitive
axis (say, the $x$-axis) can be obtained by shifting the multipole
moments of the test mass along the axis by $dX'$. This method was
used by Speake to obtain the $z$-component of force for STEP test
masses \cite{Shiomi2001}. A detailed description of the expression
for STEP is given in section 3.2 of \cite{Shiomi2002}.

Using the formula by D'Urso and Adelberger (equation (10) of
\cite{DUrso}), the leading order term of the shifted multipole
moments is given by:
\begin{equation}
\widetilde{q}_{LM} = \mp \frac{1}{2} \sqrt{\frac{(2l+3)(l \pm m+1)(l
\pm m+2)}{2l+1}}q_{lm} dX'\label{eq:shift}
\end{equation}
for $L$ = $l$+1 and $M$ = $m \pm$1. With equation (\ref{eq:V}), the
$x$-component of the force is given by:
\begin{eqnarray}
F_x = -4 \pi G \sum_{l=0}^{\infty} \sum_{m=-l}^{l} \left\{
\sqrt{\frac{(l+m+1)(l+m+2)}{4(2l+1)(2l+3)}}q_{lm}Q_{l+1,m+1}-
\sqrt{\frac{(l-m+1)(l-m+2)}{4(2l+1)(2l+3)}}q_{lm}Q_{l+1,m-1}
\right\} \label{eq:Fx}
\end{eqnarray}

From equation (\ref{eq:shift}), one can see that a positional
fluctuation $X_p$ of the test mass produces the first leading terms
of $q_{1m}$, which are proportional to $q_{00}X_p$. From equation
(\ref{eq:Fx}), one can see that these terms couple to $Q_{2,m \pm 1
}$ and produce unwanted acceleration. It should be noted that the
magnitude of the unwanted acceleration is independent of the shapes
of the test mass to the first order, but is dependent on the mass
distribution of the spacecraft ($Q_{2,m \pm 1}$). This indicates
that the mass distribution of the spacecraft has to be designed
carefully.

We apply these formulae to estimate gravitational acceleration of
the ASTROD I test mass.

\section{Monte Carlo integration}\label{st:MC}
The $x$-component of force of a test mass can also be obtained by
calculating the following formula:
\begin{equation}
F_x = G \int_{v_s}dx^3\int_{v_t}dx'^3 \frac{\rho_t (\mathbf{x}')
\rho_s (\mathbf{x}) (x-x')}{|\mathbf{x}-\mathbf{x'}|^3}
\end{equation}
We have carried out the integration over the volumes by the Monte
Carlo method \cite{Press2002}. A pair of random points, one is in
the region of $v_t$ and the other is in the region of a box that
includes $v_s$, was generated at least 10$^{8}$ times in each
simulation. The integrand was estimated and added up every time when
the random point generated in the box was inside of the region of
$v_s$. The same simulation was repeated at least 70 times. The
average and the standard deviation of the results were estimated.

\section{Cylindrical Spacecraft and the ASTROD I test
mass}\label{st:SC-TM}

We consider the gravitational acceleration between the ASTROD I test
mass, located at the centre of the spacecraft, and the spacecraft.
The test mass is a rectangular parallelepiped. Assuming a uniform
density for the test mass, the following terms of gravitational
multipole moments of the test mass (with the origin at the centre of
mass of the test mass) are null because of the geometrical
symmetries: $l$-odd terms, $m$-odd terms and terms with $m$ =
2,6,10,14,.... Therefore, the leading terms of the gravitational
multipole moments of the test mass are $q_{00}$, $q_{20}$, $q_{40}$,
$q_{4,\pm4}$ and so on. When there is a positional fluctuation of
$X_p$ along the sensitive axis, $q_{00}$ and $q_{20}$ produce the
leading terms $\widetilde{q}_{1,\pm1}$ and $\widetilde{q}_{3,\pm1}$,
respectively. From equation (\ref{eq:Fx}), $\widetilde{q}_{1,\pm1}$
couple to $Q_{20}$ and $Q_{2,\pm2}$, and $\widetilde{q}_{3,\pm1}$
couple to $Q_{40}$ and $Q_{4,\pm2}$. $Q_{2,\pm2}$ and $Q_{4,\pm2}$
are zero for homogeneous cylindrical bodies (with the origin at
their centre of mass) because of its geometrical symmetry.
Therefore, the acceleration of the test mass is given by
\begin{equation}
a_x = \frac{8 \pi
G}{M_{TM}}\left\{\frac{1}{\sqrt{30}}q_{11}Q_{20}+\frac{1}{\sqrt{21}}q_{31}Q_{40}+....\right\}
\label{eq:ax}
\end{equation}
where the relations, $q_{11}=-q_{1,-1}$ and $q_{31}=-q_{3,-1}$, are
used. By substituting the following relations, obtained from
equation (\ref{eq:shift}), into equation (\ref{eq:ax}),
\begin{eqnarray}
q_{11}=-\sqrt{\frac{3}{2}} q_{10} X_p, \;\; \;\; \;
q_{31}=-\sqrt{\frac{21}{5}} q_{20} X_p
\end{eqnarray}
we obtain,
\begin{equation}
a_x = -\frac{4 \pi G}{M_{TM}} \sqrt{\frac{1}{5}}
\{q_{00}Q_{20}+2q_{20}Q_{40}+...\}X_p \equiv -K_x X_p
\label{eq:ax1}
\end{equation}
where $K_x$ is a coupling constant.

We assume that the outer dimensions of the spacecraft is 2.0 m long
by a diameter of 2.5 m and the thickness is 5 mm, and it has a
uniform density of 2300 kg m$^{-3}$; the mass of the spacecraft is
292 kg. For this spacecraft, $Q_{20}$ = 3.31 kg m$^{-3}$ and
$Q_{40}$ = 4.07 kg m$^{-5}$. For the ASTROD I test mass, $q_{00} =
\frac{M_{TM}}{\sqrt{4 \pi}}$ = 0.493 kg and $q_{20}$ = $-$1.17
$\times$ 10$^{-4}$ kg m$^{2}$. Therefore, from equation
(\ref{eq:ax1}), $K_x \approx$ 3.5 $\times$ 10$^{-10}$ s$^{-2}$. The
first term of equation (\ref{eq:ax1}) is larger than the second term
by more than three orders of magnitude. Therefore, in order to
reduce the magnitude of this gravitational coupling, $Q_{20}$ has to
be minimized.

To shorten the simulation time and decrease the uncertainty, we have
carried out the Monte Carlo integration for a smaller cylinder (its
inner radius, outer radius, outer length and thickness of each end
disk are 45 mm, 50 mm, 100 mm and 26 mm, respectively). The density
of the cylinder is assumed to be 3000 kg m$^{-3}$. 10$^{8}$ random
points were generated in the region of the test mass and the region
of a box (100 $\times$ 100 $\times$ 100 mm$^3$), in which the
cylinder can fit. The simulation was repeated 70 times. From the
average of the 70 results, we have obtained $a_{x}$ = $-$ 2.85
$\times$ 10$^{-10}$ m s$^{-2}$ when the test mass was shifted by 1
mm along the $x$-axis. The standard deviation of the 70 results was
1.4 \%. For this cylinder, $Q_{20}$ = 2.80 $\times$ 10$^{3}$ kg
m$^{-3}$ and $Q_{40}$ = 2.71 $\times$ 10$^{5}$ kg m$^{-5}$.
Therefore, using equation (\ref{eq:ax1}), we obtain $a_{x}$ = $-$
2.8 $\times$ 10$^{-10}$ m s$^{-2}$ for $X_p$ = 1 mm. The ratio of
the magnitude of the first and second terms in equation
(\ref{eq:ax1}) is $\sim$ 4.5 \% in this configuration.

\section{A rectangular-parallelepiped box and the ASTROD-I test
mass}\label{st:RP-TM}

We consider the gravitational interaction between the test mass and
a rectangular-parallelepiped box. We assume that the test mass is
enclosed in the box. The gravitational acceleration between them has
the similar relation with equation (\ref{eq:ax1}) when the $x$ and
$y$ dimensions of the box are the same; $Q_{2,\pm2}$ and
$Q_{4,\pm2}$ are also zero in this geometry.

This configuration is similar to the electrode box for the
capacitive sensing and the test mass. The gaps between the opposing
sides of the test mass and the electrode box are 2 mm. Therefore,
the inner dimensions of the electrode box are 54 $\times$ 54
$\times$ 39 mm$^{3}$. The distance between the electrode box and the
test mass is so close that we need to consider higher terms to
estimate the gravitational acceleration between them. For a
simplicity, we consider a box larger than the electrode box: the
inner dimensions are 100 $\times$ 100 $\times$ 70 mm$^3$. We assume
that the thickness of each wall of the box is 5 mm and it has a
uniform density of 10280 kg m$^{-3}$. For this box, the acceleration
of the test mass is given by
\begin{eqnarray}
a_x &=& -\frac{4 \pi G}{M_{TM}}
\left\{\frac{1}{\sqrt{5}}q_{00}Q_{20}+\frac{2}{\sqrt{5}}q_{20}Q_{40}+\frac{5}{\sqrt{13}}q_{40}Q_{60}+2\sqrt{\frac{15}{13}}q_{44}Q_{64}+...\right\}X_p
\label{eq:ax-rp-large} \\
 &\approx& -5.6 \times 10^{-8} X_p
\end{eqnarray}
where $q_{40}$=$-2.63 \times$ 10$^{-8}$ kg m$^4$, $q_{44}$ = $-1.14
\times$ 10$^{-7}$ kg m$^4$, $Q_{20}$ = 5.93 $\times$ 10$^{2}$ kg
m$^{-3}$, $Q_{40}$ = 2.28 $\times$ 10$^{5}$ kg m$^{-5}$,
$Q_{60}$=$-4.57 \times$ 10$^{8}$ kg m$^{-7}$ and $Q_{64}$ = 2.47
$\times$ 10$^{7}$ kg m$^{-7}$. In equation (\ref{eq:ax-rp-large}),
the first term is dominant. To reduce the acceleration disturbance,
$Q_{20}$ has to be minimized. We have carried out the Monte Carlo
method for this configuration. We have generated 2 $\times$ 10$^{9}$
random points in the regions of the test mass and a box whose
dimensions are identical to the outer dimensions of the box (110
$\times$ 110 $\times$ 80 mm$^3$). This simulation was repeated 70
times. From the average of the 70 results, we obtain $a_x$ = $-$
5.60 $\times$ 10$^{-11}$ m s$^{-2}$ when the test mass was shifted
by 1 mm along the $x$-axis. The standard deviation of the 70 results
was 0.7 \%.

We have carried out the Monte Carlo simulation for the electrode box
with the inner dimensions of 54 $\times$ 54 $\times$ 39 mm$^{3}$. We
assume that the thickness of each wall of the electrode box is 5 mm
and the density is 10280 kg m$^{-3}$. 2 $\times$ 10$^{9}$ random
points were generated in the region of the test mass and the region
of a box whose dimensions are identical to the outer dimensions of
the electrode box (64 $\times$ 64 $\times$ 49 mm$^{3}$). This
simulation was repeated 90 times and the average of the 90 results
has resulted in $a_x$ = $-$ 3.93 $\times$ 10$^{-11}$ m s$^{-2}$ when
the test mass was shifted by 1 mm along the $x$-axis. The standard
deviation of the 90 results was 1.9 \% of the average. This result
of the simulations indicates that the magnitude of the gravitational
stiffness is $\approx$ 4 $\times$ 10$^{-8}$ s$^{-2}$. This is
approximately same as the value of gravitational stiffness we
currently use for the estimation of acceleration disturbances in
ASTROD I and as the gravity gradient uncertainty used in current
error estimates for LISA spurious accelerations \cite{Stebbins2004}.
For this configuration, contributions from the leading terms in
equation (\ref{eq:ax-rp-large}) are comparable. To reduce the
gravitational coupling, each moment has to be minimized.

\section{Discussion}

We have estimated gravitational acceleration between the ASTROD I
test mass and cylindrical bodies, and rectangular parallelepiped
bodies by using the analytical method and the Monte Carlo method.
The results obtained by the two methods were consistent.

From the estimations described above, one of the effective ways to
reduce the gravitational couplings seems to reduce the magnitude of
$Q_{20}$. This can be done by choosing the geometries of
constructing materials of the spacecraft. For a hollow cylindrical
body with an inner radius $A$, outer radius $B$, outer length $L$
and thickness of the end disks $T$, $Q_{20}$ is zero when they
satisfy the relation: $\frac{A}{B}=1-2\frac{T}{L}$. For the
spacecraft we have considered in section \ref{st:SC-TM}, when the
thickness is 4 mm, the dominant term, proportional to
$q_{00}Q_{20}$, becomes zero. For a box, $Q_{20}$ is null when it is
a cube.

For constructing materials closer to the test mass, contributions
from higher terms are comparable and they have to be reduced
simultaneously. In practice, many parts of the mass distribution of
the spacecraft are dictated by technical requirements. Therefore,
the most practical way of reducing the magnitude of higher terms
might be to have a test-mass shape design that minimizes the
magnitude of $q_{lm}$. In this regard, one of the ideal shapes for
the test mass is a sphere with homogeneous density; it only has
$q_{00}$. If $Q_{20}$ of materials close to the spherical test mass
is designed to be zero, the gravitational stiffness would be reduced
significantly. Other more realistic favoured shapes for the test
mass may be a cube or a cylinder with the aspect ratio of $l/a$ =
$\sqrt{3}/2$, where $a$ and $l$ are the radius and half-length of
the cylinder; their $q_{20}$ is zero.

In summary, it can be said that, in order to minimize the
gravitational couplings, the spacecraft mass distribution, $Q_{2,m
\pm 1}$, and higher gravitational moments of the test mass
($q_{lm}$) have to be minimized. However, the design of the test
mass and spacecraft will be a compromise between the requirements
from the shape designs to minimize the gravitational couplings and
other technical requirements. Therefore, an elaborate gravitational
modelling for the final design of the test mass and spacecraft mass
distribution will be necessary to ensure that the resultant
gravitational coupling is sufficiently small. It might be impossible
 to model all the mass distributions on board precisely;
 complicated shaped objects are difficult to model and some approximation might be necessary in the modelling.
 This limit on modelling needs to be considered in the estimation of
uncertainty in the gravity gradients effects. If the resultant
gravitational coupling were found too large, it needs to be adjusted
by employing compensation mass to reduce the local gravity
gradients. The analytical method provides the knowledge of which
gravitational moments contribute significantly to the resultant
gravitational coupling. It would help to figure out the optimal
designs for the test mass and the constructing materials, and the
configuration of mass compensation if necessary.

In practice, even if the test mass and the spacecraft mass
distribution are designed perfectly to make the gravitational
coupling sufficiently small, some imperfections, such as machining
tolerances and density inhomogeneities, in the test mass and the
composing materials of the spacecraft, could cause unwanted
gravitational couplings. These practical issues also have to be
studied carefully.

\section{Conclusions}
We have carried out gravitational modelling for the current test
mass design and simplified configurations of ASTROD I by using the
analytical method and the Monte Carlo simulation. In order to
minimize the gravitational couplings, the spacecraft mass
distribution, $Q_{2,m \pm 1}$, and higher gravitational moments of
the test mass ($q_{lm}$) have to be minimized. The designs of the
test mass and spacecraft mass distribution will be a compromise
between the requirements from the shape designs to minimize the
gravitational couplings and other technical requirements. To ensure
that the resultant gravitational coupling is sufficiently small, an
elaborate gravitational modelling for the final design of the test
mass and spacecraft mass distribution is necessary. Our analyses can
be applied to figure out the optimal designs of the test mass and
the constructing materials of the spacecraft, and the configuration
of compensation mass to reduce local gravity gradients.

\section*{Acknowledgements}
The author thanks W.-T. Ni for his useful comments on this work and
the manuscript, and A. Pulido Pat\'{o}n and H. Hatanaka for their
comments on the manuscript. This work was funded by the National
Science Council.


\section*{References}
\begin{thebibliography}{00}
\bibitem{ASTROD1} A. Bec-Borsenberger {\it et al} 2000 {\it ASTROD} ESA F2/F3 Mission Proposal; and references therein
\bibitem{ASTROD2} W.-T. Ni 2002 {\it Int. J. Mod. Phys.} {\bf D11} 947; and references therein
\bibitem{Ni2002} W.-T. Ni {\it et al} 2002 {\it Int. J. Mod. Phys.} {\bf D11} 1035; and references therein
\bibitem{Tang2004} Y. Xia, W.-T. Ni, C.-J. Tang, and G. Li 2006 Orbit Design and Orbit Simulation for ASTROD I,
{\it Gen. Relat. Gravit.} {\bf 37}, in press
\bibitem{Shiomi2005} S. Shiomi and W.-T. Ni 2005 Acceleration disturbances and requirements for ASTROD
I, paper in preparation
\bibitem{Xu2003} X. Xu and W.-T. Ni 2003 {\it Adv. Space Res.} {\bf 32}
No. 7 1143-1146
\bibitem{STEP} http://einstein.stanford.edu/STEP/
\bibitem{Lockerbie1993} N. A. Lockerbie, A. V. Veryaskin and X.
Xu 1993 {\it Class. Quantum Grav.} {\bf 10} 2419
\bibitem{Ni-Shiomi2004} W.-T. Ni, S. Shiomi and A.-C. Liao 2004 {\it Class. Quantum Grav.} {\bf 21} 641
\bibitem{Ni2004}  W.-T. Ni {\it et al} 2003
ASTROD I: Mission Concept and Venus Flybys, {\it Proc.5th IAA Intel
Conf. On Low-Cost Planetary Missions, ESTEC, Noordwijk, The
Netherlands, 24-26 September 2003}, ESA SP-542, pp. 79-86, November
2003; {\it Acta Astronaut.} in press, 2006
\bibitem{Shiomi2001} S. Shiomi, R. S. Davis, C. C. Speake, D. K. Gill and J. Mester 2001 {\it Class. Quantum
Grav.} {\bf 18} 2533
\bibitem{Shiomi2002} S. Shiomi 2002 Test mass metrology for tests of the Equivalence
Principle, {\it Ph.D. thesis}, University of Birmingham, Edgbaston,
Birmingham UK
\bibitem{DUrso} C. D'Urso and E. G. Adelberger 1997 {\it Phys. Rev.
D} {\bf 55} 7970
\bibitem{Press2002} W. H. Press, S. A. Teukolsky, W. T. Vetterling
and B. P. Flannery 2002 {\it Numerical Recipes in C++ The art of
Scientific Computing Second Edition}, Cambridge University Press,
section 7.6
\bibitem{Stebbins2004} R. T. Stebbins {\it et al} 2004 {\it Class. Quantum Grav.} {\bf 21}
S653
\end{thebibliography}
\end{document}